\documentclass[referee]{aa} 
\usepackage{graphicx,natbib,longtable}
\usepackage[figuresright]{rotating}
\begin{document}
   \title{Time series photometry of the dwarf planet ERIS (2003 UB313) }

   \author{Giovanni Carraro
          \inst{1,2}
          \and
          Michele Maris\inst{3}
          \and
          Daniel Bertin\inst{2}
          \and
          M. Gabriela Parisi\inst{2,4}
          }

   \offprints{G. Carraro}

\institute{
              $^1$Dipartimento di Astronomia, Universit\`a di Padova,
                 Vicolo Osservatorio 2, I-35122 Padova, Italy\\
              $^2$Departamento de Astronom\'ia, Universidad de Chile,
                  Casilla 36-D, Santiago, Chile\\
              $^3$INAF, Osservatorio Astronomico di Trieste, Via
                  Tiepolo 11, I-40013 Trieste, Italy\\
              $^4$Facultad de Ciencias Astron\'omicas y Geof\'{\i}sicas de la
                 UNLP, IALP-CONICET, Paseo del Bosque s/n, La Plata,
                 Argentina\\}

   \date{\today}


  \abstract
   {
   The dwarf planet Eris (2003 UB313, formerly known also as ``Xena'') is the largest
   KBO up to now discovered. Despite being larger than Pluto and bearing many
   similarities with it, it has not been possible insofar to detect any significant
   variability in its light curve, preventing the determination of its period and axial ratio.
  }
   {
   We attempt to assess the level of variability of the Eris light curve by
  determining its BVRI photometry with a target accuracy of 0.03~mag/frame in R
  and a comparable or better stability in the calibration.
  }
   {Eris has been observed between November $30^\mathrm{th}$ and
    December $5^\mathrm{th}$ 2005 with the Y4KCam on-board the 1.0m Yale telescope at
    Cerro Tololo Interamerican Observatory, Chile in photometric nights.
  }
   {
   We obtain 7 measures in B, 23 in V, 62 in R and 20 in I.
   Averaged B, V, and I magnitudes as colors
   are in agreement within $\approx 0.03$~mag
   with measures from Rabinowitz et~al. (2006)
   taken in the same nights.
   Night-averaged magnitudes in R shows a statistically significant
   variability over a range of about $0.05\pm0.01$~mag.
   This can not be explained by known systematics, background objects
   or some periodical variation with periods less than two days in the
   light-curve.
   The same applies to  B, V and to less extent to I
   due to larger errors.
   }
   {
   In analogy with Pluto and if confirmed by future observations,
   this ``long term'' variability
   might be ascribed to a slow rotation of Eris, with periods longer than 5~days,
   or to the effect of its unresolved satellite ``Dysnomea'' which may contribute
   for $\approx0.02$~mag to the total brightness.
   }


   \maketitle
%

\section{Introduction}

Since its discovery
the dwarf planet 2003 UB313 has attracted a lot of attention
being the first Trans Neptunian Object (TNO) larger than Pluto ever
discovered \citep{brown:2005}.
This object, recently baptized ``Eris''
\citep{IAU},
revealed a number of features in common with
Pluto, despite being a member of the family of the scattered TNO
 \citep{sheppard:2006}.

As an example, like Pluto Eris has a satellite named ``Dysnomea''
with orbital period of about two weeks,
a brightness of about $2\%$ of that of Eris,
a semi-major axis of
$\approx 5\times10^4$~Km \citep{brown:2006a}.
Eris IR spectrum is clearly dominated by CH$_4$ absorption bands
\citep{brown:2006b} and perhaps shows $N_2$ bands
 \citep{Licandro:etal:2006}.
When compared with other
TNOs its colors are quite neutral and not significantly reddened
\citep{rabinowitz:2006}.
Its phase function
at small phase angles is quite flat
\citep{rabinowitz:2006}.
Up to now its light curve did not
reveal any trace of significant variability or periodicity
\citep{brown:2005,rabinowitz:2006}.
These features suggests Eris to be an icy body which is subject to
frequent resurfacing likely due to evaporation and redeposition of a
tiny atmosphere as its heliocentric distance changes
\citep{brown:2005,rabinowitz:2006}.
\\
In this Letter we present BVRI photometry of Eris obtained during 5
nights in late 2005 with the aim of building up a light curve
and searching for possible periodicity. The same data-set is used to
better constrain the optical colors of the object.


 \begin{table*}
 \centering
 \begin{tabular}{cccccc}
 \hline\hline
   &      & B & V & R & I \\
 Night & Date & [mag]& [mag]& [mag]& [mag] \\
 \hline
 1 & Nov. 30, 2005 & $19.619 \pm 0.041$ & $18.766 \pm 0.025$ & $18.384 \pm 0.010$ & $18.019 \pm 0.030$ \\
 2 & Dec.  1, 2005 & $19.540 \pm 0.038$ & $18.768 \pm 0.015$ & $18.368 \pm 0.007$ & $18.029 \pm 0.032$ \\
 3 & Dec.  2, 2005 & $19.651 \pm 0.066$ & $18.772 \pm 0.015$ & $18.388 \pm 0.007$ & $17.948 \pm 0.022$ \\
 4 & Dec.  3, 2005 & $19.678 \pm 0.052$ & $18.788 \pm 0.016$ & $18.397 \pm 0.007$ & $17.975 \pm 0.022$ \\
 5 & Dec.  4, 2005 & $19.616 \pm 0.077$ & $18.802 \pm 0.015$ & $18.422 \pm 0.007$ & $18.039 \pm 0.027$ \\
 \hline\hline
 \end{tabular}
 \caption{
  Weighted averages of BVRI for each night.
  \label{tab:avermag}}
 \end{table*}

   \begin{figure*}
   \centering
   \resizebox{\hsize}{!}{
   \includegraphics{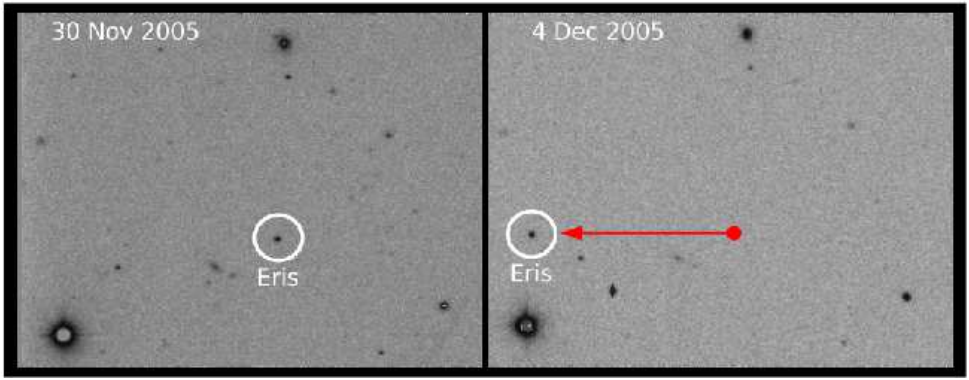}}
   \caption{ Eris position in the first (left panel) and last night
   (right panel). The area is 3.5 $\times$ 2.7 arcmin. North in down,
   East to the right. The arrow indicates Eris motion over the whole run.
          \label{fig:map}}
    \end{figure*}

\section{Observations and Data Reduction}
We observed Eris over 5 consecutive nights (November 30 to
December 4, 2005). Photometric data were obtained with the
Y4KCam CCD on-board the Yale 1.0m telescope at Cerro Tololo
Interamerican Observatory, which is operated by the SMARTS
consortium\footnote{http://www.astro.yale.edu/smarts/}.
The Y4KCam instrument is a 4096 $\times$
4096 CCD with a pixel scale of 0.289$^{\prime\prime}$, which allows
one to observe a field 20 arcmin on a side on the sky.
An image of the field around Eris is shown in Fig.~\ref{fig:map}.
A series of images in BVRI was acquired in order to constraint both
the light curve and the colors. A total of 7 images in B, 23 in V, 62
in R and 20 in I has been obtained over the observing run, and the exposure
times were 300-600 secs.
The nights were all photometric but for the last one (December 4, 2005),
with typical seeing ranging from 0.8 to 1.2
arcsec. The images were cleaned and pre-reduced using the
pipeline developed by Phil
Massey\footnote{http://www.lowell.edu/users/massey/obins/y4kcamred.html}.
To extract Eris magnitudes we used the {\it QPHOT} task within
IRAF\footnote{IRAF
is distributed by the National Optical Astronomy Observatories,
which are operated by the Association of Universities for Research
in Astronomy, Inc., under cooperative agreement with the National
Science Foundation.}, which allows one to measure aperture
photometry. For Eris we used a small aperture (7 pixels).
Together
with Eris we measured 5 field stars with roughly the same magnitude
($17.53 \leq  \langle R \rangle \leq 19.25$).
Due to the slow motion of Eris and the wide field covered by the CCD, we could
measure the same 5 stars every night
and thus tie the photometry to the same
zero point for  the entire data-set. For the field stars we used a
larger aperture (18 pixels).
Absolute magnitudes
were derived by shifting Eris magnitudes to the first night using the
reference field stars. A set of bright stars in the first night
were used to
aperture-correct the magnitudes. Aperture corrections were found to
be small, of the order of 0.05-0.12 mag. The zero points
of the photometry was then obtained through the observation of 50 standard
stars in the Landolt (1982) fields PG0231, SA92 and Rubin149,
using as well a large aperture of 18 pixels.
The magnitudes were also color-corrected using Eris mean colors.
The final photometry, consisting of 115 data points, is
reported in Tab.~2 together with the photometric error, UT time,
and filter.
In the same period we observed, Rabinowitz et al. (2006) have obtained
2 B, 4 V and 2 I\,  images of Eris. We compared our photometry with
their one, and found  a good agreement, being $\Delta B = 0.05$,
$\Delta V = -0.032$ and $\Delta I = 0-0.022$, in the sense our
photometry minus their one. We have not direct
comparison with R, since these authors did not publish data in R
for these nights.

\section{Light curve and period hunting}

\def\DeltaR{\Delta R}

\begin{figure}
   \centering
   \resizebox{\hsize}{!}{
   \rotatebox{-90}{
   \includegraphics[height=17cm]{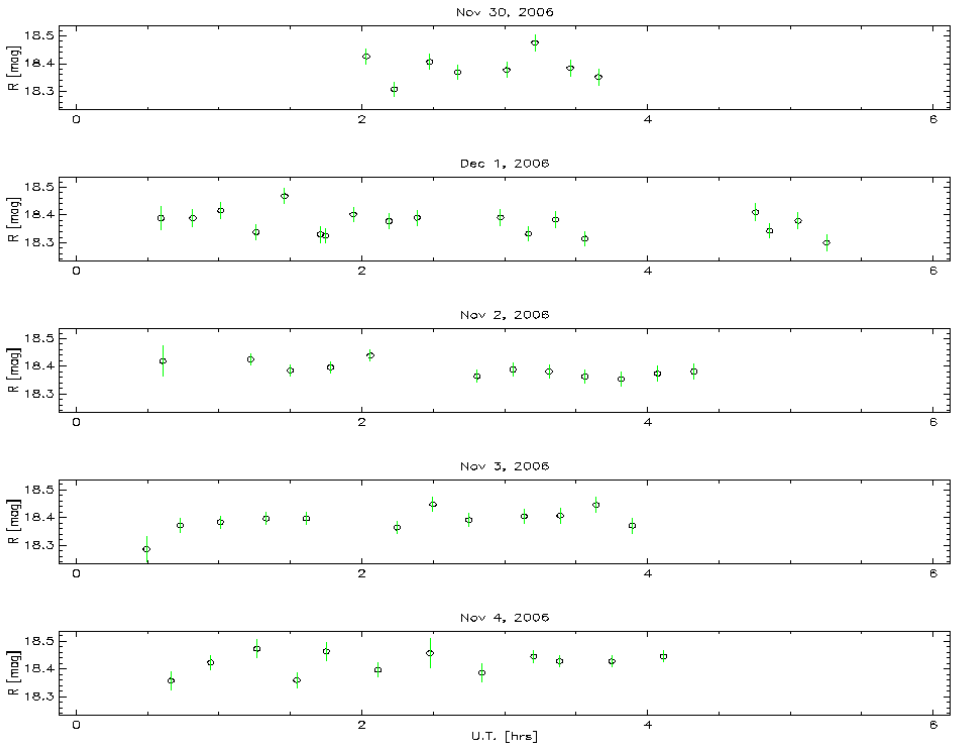}}}
   \caption{R Light curves of Eris (2003 UB313). From the top to the bottom
          the light curve for Nov. 30 to Dec. 4, 2005  are shown.
          \label{fig:lc}}
    \end{figure}

\begin{figure}
   \centering
   \resizebox{\hsize}{!}{
   \includegraphics{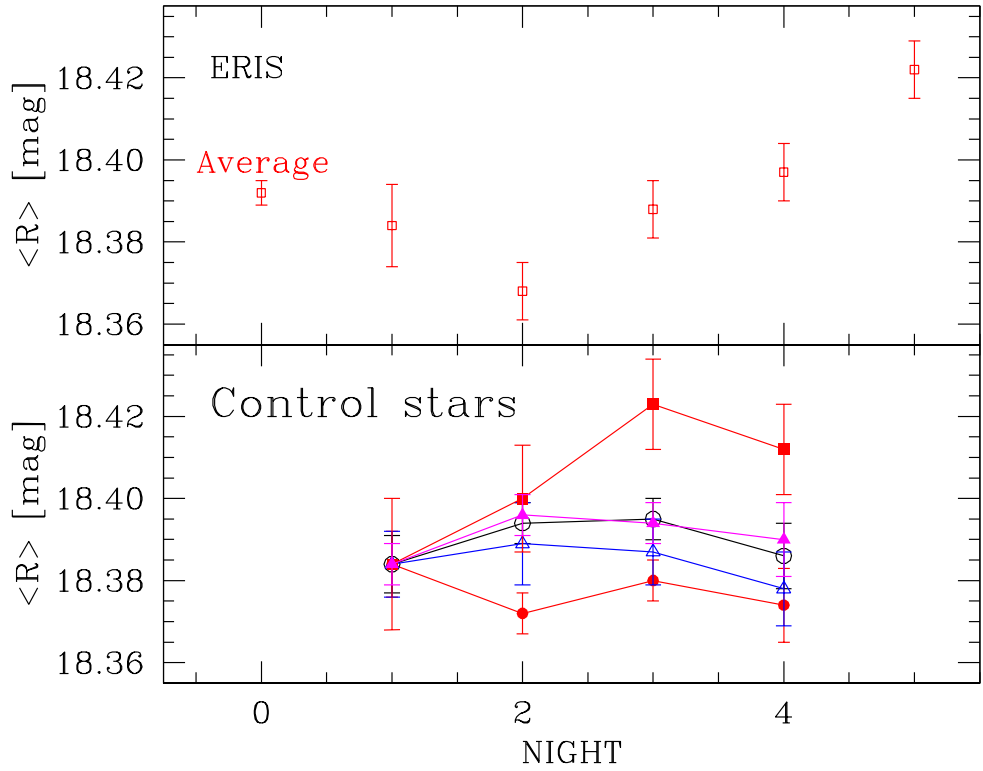}}
   \caption{{\bf Upper Panel:}
         Eris weighted averages R mag as a function of night (see Tab.~\ref{tab:avermag}).
          The point in blue (dark gray in bw print) at night 0 is the weighted
          average over the 5 nights.{\bf Lower panel:} Weighted
         averages of R as a function of the night for the 5 field
         stars discussed in the text. Only the data for the first 4
         nights are plotted, since the fifth was not photometric.
         The magnitude of these stars have been arbitrarily shifted to
         the first night Eris averarge R mag.
          \label{fig:avermag}}
    \end{figure}

Fig.~\ref{fig:lc} shows night-by-night time dependencies in the R band.
It is evident that, even removing the four measures with anomalously large errors
($\DeltaR > 0.04$~mag), the dispersion of data can not be attributed just to 
random errors. 
The weighted average for all of the five nights gives $\bar R = 18.3916 \pm 0.0033$~mag
with $\chi^2 = 136$ and 57 d.o.f. (degree of freedom),
a $\chi^2$-test rejects the hypothesis of random fluctuations
at a confidence level ($\mathrm{c.l.}$) $\sim 2\times10^{-8}$
insensitive to the  exclusion of the
bad measures.
Inspection of the R frames shows that Eris is moving very slowly
in an uncrowded field (see Fig.~\ref{fig:map}), with no evident objects in background.
Moreover, due to the short exposures and the low proper motion, both Eris and field stars
are round without  trailing.
All together this seems to exclude at least the most common systematic effects.
The observing conditions (heliocentric distance $r_{\mathrm{h}}$, geocentric distance
$\Delta$, as phase angles $\alpha$)
could be responsible for the effect.
However, during the five nights they were fairly stable. In fact
the object moves of about 95~arcsec in 5 nights.
The change in $r_{\mathrm{h}}\Delta$ explains
no more than $1.7\times10^{-3}$~mag.
On the other hand, the phase angle changes of $\Delta \alpha = 0.0246^\circ$ during the observations.
No phase coefficients in R have been published so far for Eris,
but assuming as an upper limit the same phase coefficient of V in
\cite{rabinowitz:2006},
the phase effect would account at most for $3.5\times10^{-3}$~mag.
In conclusion,  obvious changes in the observing conditions excludes geometrical
effects.\\
Tab.~\ref{tab:avermag} reports weighted averages for magnitudes taken in the same
night, while the lower frame of Fig.~\ref{fig:avermag} displays the same data for the R filter.
A clear trend appears in R for nights 2 to 5.\\
A $\chi^2$ test rejects the hypothesis of random fluctuations
at c.l.~$\approx 2\times10^{-6}$,
inclusion of night 1 does not change this result.
This is robust against selection of data according to the U.T.
of observation (as evident from Fig.~\ref{fig:lc}
in night 1 Eris has been observed just between
$U.T. = 2$ and $U.T. = 4$ considering only data in that U.T. interval
does not change the result) and replacing weighted averages with median estimation
of nightly centroids.
The difference between nights 5 and 2 is
$0.054\pm0.010$~mag, equivalent to $5.4\sigma$.
A linear fit for nights 2 to 5 gives a slope $R' \approx 0.0170 \pm 0.0002$~mag/day
with $\chi^2_{\mathrm{linear}}=0.99$\ equivalent to c.l.~$\approx0.6$ that residual fluctuations are just
due to errors.
A parabolic fit including all the nights gives $\chi^2_{\mathrm{parabolic}} = 2.35$\ equivalent to a
c.l.~$\approx0.3$.\\
 %
%
 %
{\bf The lower panel of Fig.~\ref{fig:avermag} compares variations for 5 field stars having R in the 
approximated range 
$17.5$~mag - $19.3$~mag encompassing the range of Eris R magnitudes. 
 Magnitudes are measured frame by frame and averaged over each night in the same manner of Eris data.
 To highlight the variations, the first night of each serie has been
 shifted to the averaged R for Eris,  R=18.39~mag.
 It is evident that field stars are stable with peak-to-peak variations in R of about 0.01~mag. 
 The only star departing from this value is the weakest in the serie having $R=19.3$~mag. 
 In addition the expected random errors for field stars are similar to the random errors for Eris. Larger
 errors appearing for R larger then $19$~mag.
 A convincing test of the calibration stability comes from the fact that  
 variability indicators for field stars 
(either peak-to-peak variation, the {\it r.m.s.} between the 4 nights, 
the $\chi^2$ for fitting against a constant value or better the related significativity) 
plotted as a function of their mean magnitude are constant for R up to $\approx19$~mag.
Moreover, for Eris the indicators of variability allways differ significantly 
from the values obtained for field stars below $R = 19$~mag. 
Peak-to-peak variations for field stars is $\approx 0.011$~mag - $0.012$~mag v.z. Eris 0.029~mag. 
Night-by-night rms for the field stars is $\approx0.005$ - $0.006$~mag,
v.z. Eris 0.012~mag. 
The significativity of fluctuations for field stars is always below the 60\% level v.z. 
Eris showing fluctuations with a significativity larger than 98\%.
In addition, field starts fluctuations are not very much correlated with Eris fluctuations,
in some cases field stars are anticorrelated with Eris and correlations are not much significant.
 All this supports the idea that Eris brightness variations are not due to calibration errors.}
%
 %
Looking at the other filters the same trend in nights 2 to 5 appears
for V, B and marginally I but with a lower significance owing to larger errors.
We excluded that the trend is connected to fluctuations in the zero point calibration
as derived from standard stars.
The night-by-night zero point for R, $R_0$,
is spread of $\Delta R_0 \approx 0.004$~mag consistent with its {\it r.m.s.}
$\sigma_{R_0}\approx0.008$~mag and has just a marginal
trend with slope $4\times10^{-4}$~mag/day, to be compared with the spread of Eris
over the first four nights of $\Delta R_{\mathrm{Eris}} = 0.029$~mag.
For V, $\Delta V_0 \pm \sigma_{V_0} \approx 0.005 \pm 0.01$~mag
to be compared with
$\Delta V_{\mathrm{Eris}} = 0.022$~mag.
Besides, V and R are correlated the
correlation coefficient $\rho_{\mathrm{VR}} = 0.92$.
At the same time V-R computed night-by-night is fairly stable. A fit against the
case of constant V-R has
$\chi^2 = 0.85$ corresponding to a c.l.~$\approx0.93$ that fluctuations about the
averaged value ($V-R = 0.388 \pm 0.008$ from  Tab.~\ref{tab:avermag}) are just due to chance.
As a comparison the $V_0 - R_0$ on a night-by-night basis
has an r.m.s.~$=0.006$~mag with a c.l.~$\approx0.99$ for random fluctuations.
Correlation between colors in light curves is expected if
Eris is an icy body frequently resurfaced by
atmospheric freezing. In this case
a uniform layer of frozen gasses should hide color variations.\\
B and I have less precise calibration and random errors and sparser coverages,
but for completeness it is worth to extend the discussion to these data too.
B and I are less correlated with R having respectively
$\rho_{\mathrm{BR}} = 0.465$,
$\rho_{\mathrm{IR}} = 0.137$.
The correlation between B and R is very sensitive to the exclusion of
the last night. Then for the first four nights
$\rho_{\mathrm{BR}} = 0.995$,
In addition the c.l. against random fluctuations are just 0.23 and 0.04 respectively
for B and I.
Again,  the variability in B and I can not be reconciled with variations in $B_0$ and
$I_0$ since
$\Delta B_{\mathrm{Eris}} = 0.138$~mag
and
$\Delta I_{\mathrm{Eris}} = 0.081$~mag
while $\Delta B_0 \pm \sigma_{B_0} \approx 0.03 \pm 0.01$~mag and
$\Delta I_0 \pm \sigma_{I_0} \approx 0.039 \pm 0.018$~mag.
Note the different behavior of I in the second night.
While B, V, R in night two have lower or equal magnitudes respect to night one and
three, I shows the opposite trend.
Indeed after removing the second night
$\rho_{\mathrm{IR}} = 0.57$, while removing even the first $\rho_{\mathrm{IR}}=0.998$.\\
If this tiny time dependence is not due to some unaccounted problem in the data,
would be this a sign of an aliasing with short term variability?
We attempt to assess whether the dispersion in the data can be ascribed to some periodical
variations in the light curve over periods shorter than 5 nights.
The phase dispersion minimization method applied to data binned in chunks of 1~hour,
favors periodicity of about 30~hrs.
Other possible periods are much sensitive to the details of the method, as the
number of phase bin or the step in periods.
An inspection of data plotted as a function of phase for a 30~hrs periods
suggests a scattered, non sinusoidal light curve, with maximum peak-to-peak
variation of $\approx 0.06$~mag and a single maximum, but
the fitting is marginal ($\chi^2 = 116.74$ with c.l.~$=2\times10^{-5}$ that deviations from the
fit are not just due to random errors).
The periodogram of data does not allow us to identify any
noticeable periodicity between 1 and 100 hours.
This is true even after exclusion of periods heavily affected by aliasing
(6~hrs, 8~hrs, 12~hrs, 24~hrs and 48~hrs). In particular
the 30~hrs period is just outside the 24~hrs side-lobe and the improvement in
the $\chi^2$ for fitting data with an $\approx 30$~hrs period is again
marginal.
To have a more robust test we play numerical experiments with simulated sinusoidal
signals plus noise. Here we consider periods in the range $1 - 100$~hrs, amplitudes
$\le 0.05$~mag, constant $R$ magnitude in the $\pm5\sigma$ of our data and phases in
the range $0 - 2\pi$. Simulated data has been re-binned on a night-by-night basis and
compared to night averaged data computing the corresponding $\chi^2_{\mathrm{sin}}$.
As a comparison we take $\chi^2_{\mathrm{sin}}$
with $\chi^2_{\mathrm{const}}$, $\chi^2_{\mathrm{linear}} = 2.36$
(computed over 5 nights) and $\chi^2_{\mathrm{parabolic}}$
as defined before.
Our results show no significant improvement in the fit by assuming a
sinusoidal signal in the data.
In at most $3\%$ of our $3\times10^5$
simulated realizations we obtained
$\chi^2_{\mathrm{sin}} < \chi^2_{\mathrm{const}}$.
The fraction drops to $0.05\%$ and $0.003\%$ respectively for
$\chi^2_{\mathrm{sin}} < \chi^2_{\mathrm{linear}}$ and
$\chi^2_{\mathrm{sin}} < \chi^2_{\mathrm{parabolic}}$.
To have an extreme case of non-sinusoidal signal we consider also the
case of a square wave with variable amplitude, period, phase and
duty-cycle obtaining a largely worst fit.
In conclusion the long term variability in our data can not be
explained by aliasing of an under-sampled short term variability.

\section{Colors}
We computed weighted mean colors of Eris. These are derived from
the  weighted mean of all the measures in each filter.
We obtain
$B-V=0.823\pm0.023$, $V-R=0.391\pm0.023$, $R-I=0.386\pm0.012$ and
$V-I=0.777\pm0.013$, quite in agreement with Rabinowitz et al. (2006,
Tab.~4).
Following the same vein of the discussion in this paper, we confirm that
the colors of ERIS are solar, with only B-V being marginally redder than
the Sun \citep{Hainaut:Delsanti:2002}. These colors corroborate
the  idea that Eris is an icy body.


\section{Discussion and Conclusions}
We have presented time series photometry in BVRI pass-bands of the
dwarf planet Eris (2003 UB313).\\
Looking at the data we have presented and analysed (in particular V
and R),
it is possible to say that some genuine time variability is present
with a reasonable level of confidence.
If this will be confirmed by further observations
it would indicate a light curve with a long term variability. Likely,
one with periods greater than 5 days and amplitudes $\approx 0.05$~mag.
Such a small amplitude would indicate a low axial ratio for Eris or that this body
is seen nearly pole-on from the Earth.\\
In the first case Eris would be more symmetric in shape than
other known KBOs or Pluto itself. \\
In the second case,
due to the large distance to the Sun,
Eris is pole on with respect to the Sun too.
Presently Eris is near its aphelion and
if it has an axial ratio comparable to that of Pluto,
we should expect that the maximum amplitude
of its light curve would be observed toward the epoch in which it will have an anomaly
of $\approx 90^\circ$.
However, even a change of $10^\circ$
in its orientation would produce a significant increment in the amplitude of its light
curve.\\
Interestingly enough  for the evolution of resurfacing is the fact that in case
Eris were seen pole-on at aphelion, it would have to be pole-on even at
perihelion.
Having a so large orbital ellipticity, the solar irradiation at aphelion
would be 6.6 times smaller than the irradiation at perihelion.
Depending on the details of resurfacing mechanism and atmospheric circulation, it
would not be a surprise to discover significant differences between
the  two hemispheres of Eris.
As an example, one can speculate that the region of the aphelion pole would
be more rich in volatiles than the opposite region.
If so, even the spectroscopic signature of the Eris surface will have to
show secular variations correlated with the light-curve amplitude.\\
\noindent
Finally, in trying to understand Eris light curve,
one cannot neglect that the presence of the un-resolved
satellite could distort it.
Dysnomea indeed may contribute for up to
$\approx 0.02$~mag to the time variability of brightness
with an expected orbital period of 2~weeks.
However, even assuming that the line of nodes of the orbit of the
satellite is oriented toward
the Sun, an eclipse or a transit would last for about one tenth of day, compatible with
the time scale of our observations over each night. But an eclipse or a transit would
cause a drop in brightness while our data suggest rather the opposite behavior. In
addition, an eclipse or a transit would affect only one night and not the subsequent
ones due to the small phase angle with which we are observing the system.

\begin{acknowledgements}
The work of GC was supported by {\it Fundacion Andes}.
The work of MM was partially supported by INAF FFO for free research 2006
({\em Fondo Ricerca Libera}).
The authors acknowledge the referee, David Rabinowitz, for useful suggestions.
\end{acknowledgements}

%

\end{document}